\documentclass[aip,apl,amsmath,amssymb,reprint]{revtex4-1}
\usepackage{amsmath}
\usepackage{textcomp}
\usepackage{bm}
\usepackage{xcolor}
\usepackage{subfigure}
\usepackage{graphicx}
\usepackage{graphics}
\begin{document}
	
\title{Insight into perovskite antiferroelectric phases: Landau theory and phase field study}

\author{Zhen Liu}
\email{z.liu@mfm.tu-darmstadt.de} 
\author{Bai-Xiang Xu}
\email{xu@mfm.tu-darmstadt.de} 

\affiliation{$^1$Mechanics of Functional Materials, Department of Materials Science, Technical University of Darmstadt, 64287 Darmstadt, Germany}

\date{\today}

\begin{abstract}
Understanding the appearance of commensurate and incommensurate modulations in perovskite antiferroelectrics (AFEs) is of great importance for material design and engineering. The dielectric and elastic properties of the AFE domain boundaries are lack of investigation. In this work, a novel Landau theory is proposed to understand the transformation of AFE commensurate and incommensurate phases, by considering the coupling between the oxygen octahedral tilt mode and the polar mode. The derived relationship between the modulation periodicity and temperature is in good agreement with the experimental results. Using the phase field study,  we show that the polarization is suppressed at the AFE domain boundaries, contributing to a remnant polarization and local elastic stress field in AFE incommensurate phases. 

\end{abstract}

\maketitle

Perovskite ABO$_3$ antiferroelectrics (AFEs) are the most representative AFE materials that display giant energy storage density\cite{qi2019linear,zhao2017lead}, giant electrocaloric effect\cite{peng2013giant,geng2015giant}, and giant electrostrictive property\cite{guo2011composition}. The properties of AFE materials are greatly influenced by their microstructures and domain morphologies\cite{cheng2009structural, gao2015electric}. Therefore, understanding the transformation of the AFE phase structures is vital for material design and discovery. The AFE phase was originally predicted by Kittel's phenomenological model that two interpenetrating sublattices have opposite polarizations\cite{kittel1951theory} like ($\uparrow\downarrow$). However, later investigations have shown that many behaviors of perovskite AFEs cannot be understood through the simple two lattice model\cite{}. Especially, lots of experiments have shown that slightly ion-doped 
perovskite AFEs generally favor AFE commensurate (AC) phases such as ($\uparrow\uparrow\downarrow\downarrow$) and ($\uparrow\uparrow\uparrow\downarrow\downarrow\downarrow$), or AFE incommensurate (AI) phases like ($\uparrow\uparrow\uparrow\uparrow\downarrow\downarrow\downarrow$) in many AFE system \cite{asada2004induced,he2005electric,ma2019uncompensated,guo2015direct}.
The modulation of the AC and AI phases is sensitive to the chemical composition and temperature, which present potential applications in domain boundary engineering. Although many theoretical models of AFEs have been proposed\cite{hatt2000landau, tagantsev2013origin, toledano2016theory}, the origin of these AFE structures is still unclear. Besides, unlike the ferroelectric domain walls that are widely studied during the past decades, the characteristic of AFE domain boundaries are lack of investigations.

It was originally proposed that the Brillouin zone-center and zone-boundary modes should exhibit softening at the AFE transformation\cite{cochran1968structure}.
However, an infrared spectra study of PbZrO$_3$ ceramics revealed only a slight softening of the zone-center modes which contribute a high dielectric constant at the Curie point\cite{ostapchuk2001polar}. A novel mechanism of AFEs is proposed that AFE transformation is driven by the softening of a single lattice mode via flexoelectric coupling\cite{tagantsev2013origin}. However, experimental measurements later show that the flexoelectric effect of PbZrO$_3$ and AgNbO$_3$ at room temperature is too small to stabilize the antiferroelectric phases\cite{vales2018flexoelectricity}. And the recent polarized IR and Raman spectroscopic study indicates that PbZrO$_3$ indeed exhibits multiple soft modes, resulting in a flat soft polarization branch rather than a local minimum near the AFE wave vector\cite{hlinka2014multiple}. On the other hand, the softening of oxygen octahedral rotational modes is known as a significant role in structural phase transformations for a variety of perovskite ferroelectrics (FE) and AFE \cite{glazer1975simple,xu1995evidence}. The complex coupling of the $\frac{2\pi}{a_0}(\frac{1}{4},\frac{1}{4}, 0)$ $\Sigma_3$ mode (antiparallel shifts of the lead ions) and the $\frac{2\pi}{a_0}(\frac{1}{2},\frac{1}{2}, \frac{1}{2})$ ($R$ point) $\Gamma_{25}$ mode (antiphase tilts of oxygen octahedron) is believed to give rise to antiferroelectricity\cite{viehland1995transmission,fthenakis2017dynamics} in PbZrO$_3$. In this work, to investigate the AFE structures, we have proposed a new phenomenological model of AFEs by considering the coupling between the polar mode and the oxygen tilt mode. The modulation of the AFE phases related to the temperature is investigated. The polarization distribution and elastic property across the domain boundaries are further calculated via the phase field simulation.

We demonstrate the new Landau free energy first in an ABO$_3$ lattice with oxygen octahedral tilt $\theta_y$ around the $y$-axis, and the A-site ions accommodate to displace in the $x$-axis. The opposite displacements of the A-site cations lead to an antiparallel polarization $p_x$ of AFE state. The simple potential can be written in the form as
\begin{equation}
\begin{split}
F=&\frac{\alpha}{2}p_x^2+\frac{\beta}{4}p_x^4+\frac{\gamma}{6}p_x^6+\frac{\sigma_0}{2}\theta_y^2(\frac{\partial p_x}{\partial z})^2+\frac{g}{2}(\frac{\partial^2 p_x}{\partial z^2})^2 \\
&+ \frac{\lambda}{2}\theta_y^2p_x^2+\Phi_\theta
\end{split}
\end{equation}
where $\alpha=\alpha_0(T-T_0)$, $T_0$ is the Curie-Weiss temperature related to the polar mode, $\sigma_0$ is a coefficient, and the constants $\alpha_0,\beta,\gamma$ $\lambda$ and $g>0$. The potential of the oxygen tilt written up to fourth-order is given by  $\Phi_\theta=\frac{k_0(T-T_\theta)}{2}\theta_y^2+\frac{b_0}{4}\theta_y^4$, in which $k_0$ and $b_0$ are positive constants, and $T_\theta$ is the transition temperature of the oxygen tilt. Thus, below $T_\theta$, the oxygen octahedral tilt angle is obtained by $\theta_y=\sqrt{-\frac{k_0(T-T_\theta)}{b_0}}$. 
The spontaneous polarization related to the polar mode can be expressed as $p_x=\eta e^{{\rm i}qz}$, where $\eta$ is the polar amplitude and $q$ is the $z$-axis polar mode waver vector. Let $\sigma=\sigma_0\theta_y^2$, then Eq.(1) can be rewritten as
\begin{equation}
F=(\frac{\alpha'}{2}+\frac{\sigma}{2}q^2+\frac{g}{2}q^4)\eta^2+ \frac{\beta}{4}\eta^4+\frac{\gamma}{6}\eta^6+\Phi_\theta
\end{equation}
where $\alpha'=\alpha_0(T-T_0+\frac{\lambda}{\alpha_0}\theta_y^2)$, which indicates that the transition temperature of the polarization can be influenced by the oxygen tilt.

The polar mode vibration is given by \cite{slonczewski1970interaction,schwenk1990phase}
\begin{equation}
\frac{\partial^2 \eta}{\partial t^2}=-\xi \frac{\delta F}{\delta \eta}
\end{equation}
where $\xi$ is a constant. Thus, we get
\begin{equation}
\omega^2(q)\propto \alpha'+\sigma q^2+gq^4
\end{equation}
where $\omega$ is the angular frequency of the optic-mode. Therefore, if $\sigma\ge0$, the soft polar mode appears at the Brillouin zone-center with $q=0$, which corresponds to the FE state. If $\sigma<0$, the minimum of the polar mode shifts away from the zone-center, which gives rise to the AI or AC phases. To make a minimum free energy of the system, the wave vector $q$ should satisfy the equation $\frac{\partial F}{\partial q}=0$, we obtain
\begin{equation}
q^2=-\frac{\sigma}{2g}=-\frac{\sigma_0}{2g}\theta_y^2 , (\sigma_0<0) \label{eq5}
\end{equation}

From Eq.(\ref{eq5}) we know that the angle of the oxygen octahedral tilt directly determines the polarization phases. If $\theta_y$ is very small that makes $q$ approximate to zero, the polarization phase is the FE incommensurate (FI) phase. If $q$ is close to the zone-boundary, the AI phase occurs. For the cases that $q$ equals $\frac{\pi}{a_0}$ or $\frac{\pi}{2a_0}$ ($a_0$ is the lattice parameter), the corresponding polarization phases are ($\uparrow\downarrow$) or ($\uparrow\uparrow\downarrow\downarrow$) AC phases, respectively. Experimentally, it has been found that the oxygen tilt temperature $T_\theta$ is far above the AFE transformation point in many perovskite AFEs, such as PbZrO$_3$ \cite{viehland1995transmission} and NaTaO$_3$\cite{rechav1994local}, indicating that the AFE transformation occurs within an already titled oxygen framework.  No anomaly in the rotation angle was observed near the AFE transition, rather the angle slowly increases with decreasing temperature. Therefore, if the oxygen tilt angle has a relatively small value at the AFE transition point, the system is likely to undergo a phase transformation from paraelectric to FI to AI to AC. This transition behavior is confirmed by the phase diagram of the La-doped lead-based solid solution in the literature\cite{asada2004induced}.

Using equation Eq.(\ref{eq5}) and $q=\frac{2\pi}{a_0}\frac{1}{n}$ ($n$ is the number of lattice in a period), we derive the equation of the periodicity as $n=\frac{2\pi}{a_0}\sqrt{\frac{-2g}{\sigma_0\theta_y^2}}$. By substituting the value of oxygen tilt to it, the periodicity of the AI phases versus the temperature can be achieved as
\begin{equation}
n=\frac{2\pi}{a_0}\sqrt{-\frac{2b_0g}{k_0\sigma_0}}\frac{1}{\sqrt{T_\theta-T}}\propto\frac{{1}}{\sqrt{T_\theta-T}} \label{eq6}
\end{equation}

\begin{figure}
	\centering
	\includegraphics[width=2.7in]{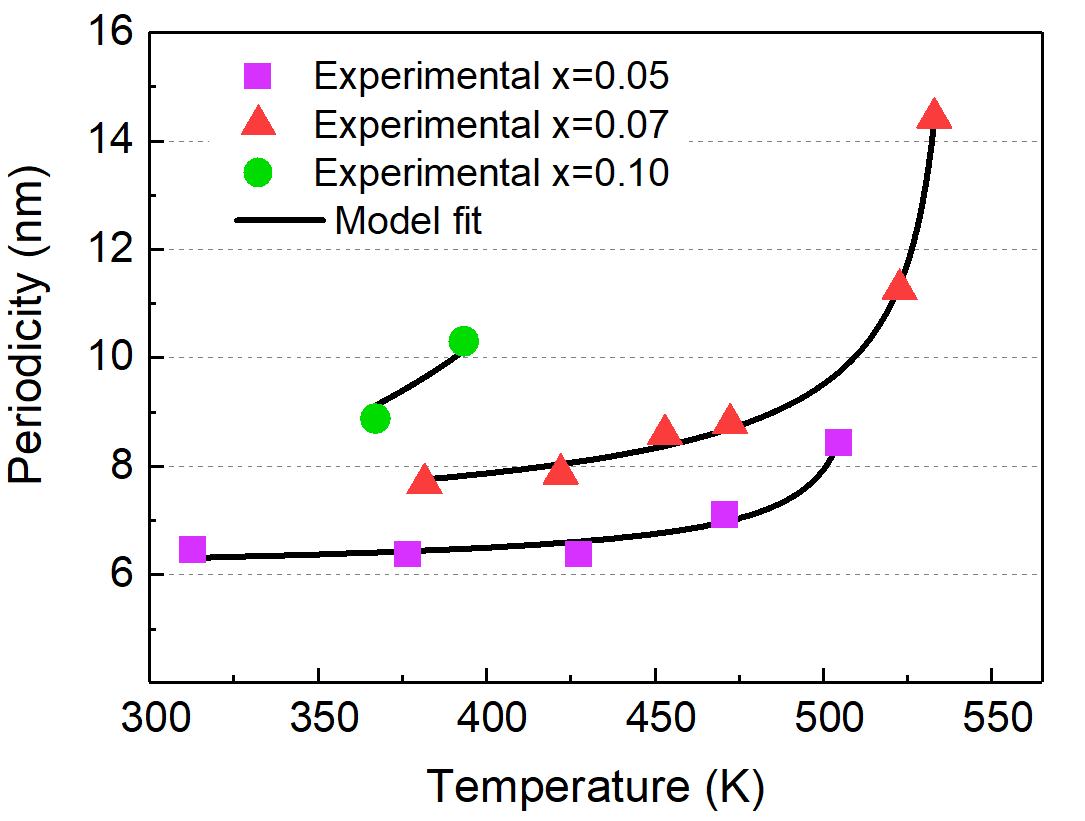}
	\caption{The periodicity of the incommensurate phases versus the temperature of the Pb(Zr$_{1-x}$Ti$_x$)O$_3$ solid solution. The square, triangle and cycle data points are extracted from literature\cite{watanabe2002features} for compositions $x=0.05, 0.07$ and $0.10$. The black curves are model fitted which are in good agreement with the experimental data. }
	\label{fig1}
\end{figure}

To examine the relationship between the periodicity and temperature of the incommensurate phases, we have compared our theoretical results to the experimental data of Pb(Zr$_{1-x}$Ti$_x$)O$_3$ (PZT) system\cite{watanabe2002features} with $x\leq 0.10$. As shown in Figure \ref{fig1}, for compositions $x=0.05$ and $x=0.07$, experimental data show the periodicity ($n$ multiply the lattice constant) decreases quickly around the AFE transformation temperature, and then descends gradually with temperature further decreases. 
The black curves are fitted based on Eq.(\ref{eq6}) using function $n=\frac{A}{\sqrt{T_\theta-T}}+C$ ($A, C$ are constants). A strong agreement with the experimental results is obtained.
 
We now develop the free energy into a three dimensional form for the perovskite system. Knowing perovskite oxides have a cubic phase  with $O_h$ symmetry above the Curie temperature, the invariant polynomials in terms of the order parameters only contain the even terms. The free energy density due to the softening of the polar mode and oxygen octahedral tilt mode, which is invariant under $O_h$ symmetry, can be written as
\begin{equation}
F=F_p+F_{p\theta}+F_{\theta}+F_{q} \label{F3d}
\end{equation}
The first term in Eq.(\ref{F3d}) is the classical free energy of the spontaneous polarization. It is given by
\begin{equation}
\begin{split}
F_p=&\sum_i(\alpha_1 p_i^2+\alpha_{11} p_i^4+\alpha_{111} p_i^6)+\alpha_{123}\prod_ip_i^2\\
&+\alpha_{12}\sum_{i\ne j}p_i^2p_j^2+\alpha_{112}\sum_{i\ne j}p_i^4p_j^2
\end{split}
\end{equation}
where $p_i$ denotes the polarization component along the three axes, $\alpha_i, \alpha_{ij} $ and $\alpha_{ijk}$ ($i, j, k$=1, 2, 3) are coefficients. The second term $F_{p
\theta}$ describes the biquadratic coupling of the polarization and oxygen octahedral tilt, which is written as $F_{p\theta}=\mu_{11}\sum_{i}p_i^2\theta_i^2+\mu_{12}\sum_{i\ne j}p_i^2\theta_j^2$, where $\theta_{i}$ denotes the component of the oxygen tilt angle, and $\mu_{ij}$ is constant. This term can be integrated into $F_p$ with modified Curie temperature related to the coefficient $\alpha_1$. The third term represents the potential of oxygen octahedral tilt, which reads $ F_\theta=\sum_{i}{(k_{1}\theta_i^2+k_{11}\theta_i^4)}+k_{12}\sum_{i\ne j}{\theta_i^2\theta_j^2} $, where $k_1=k_0(T-T_{\theta})$, $(k_0>0)$, $k_{11}$ and $k_{12}$ are assumed to be temperature independent. 

The wave vector of polar modes is determined by 
\begin{equation}
\begin{split}
F_q=&\gamma_{11}\sum_i{p_i^2 p_{i,i}^2}+\gamma_{12}\sum_{i\ne j\ne k}{p_i^2(p_{j,k}^2+p_{k,j}^2)} \\
&-\sigma_{11}\sum_i{\theta_i^2 p_{i,i}^2} - \sigma_{12}\sum_{i\ne j\ne k}{\theta_i^2 (p_{j,k}^2+p_{k,j}^2)}+F_g
\end{split}
\end{equation}
where $p_{i,j}=\frac{\partial p_i}{\partial x_j}$, and the coefficients $\gamma_{11}$, $\gamma_{12}$, $\sigma_{11}$ and $\sigma_{12}$ are positive constants. FE phase is stable if the angle of oxygen octahedral tilt is small that makes $(\gamma_{11}p_i^2-\sigma_{11}\theta_i^2)$ and $(\gamma_{12}p_i^2-\sigma_{12}\theta_i^2)$ larger than zero. Otherwise, the wave vector of the corresponding modulation will shift away from the zone-center, resulting in the appearance of IC or AFE phases. The potential $F_g$ is given in terms of the second order derivative of the polarization,
\begin{equation}
\begin{split}
F_g=&g_{11}\sum_i{(\frac{\partial^2 p_i}{\partial x_i^2})^2}+g_{12}\sum_{i\ne j}{(\frac{\partial^2 p_i}{\partial x_j^2})^2}
\end{split}
\end{equation}
where $g_{11}$ and $g_{12}$ are positive constants. 
\begin{figure} 
	\centering 
	\subfigure[ $n$=7]{ \label{fig2:a} 
		\includegraphics[width=1.4in]{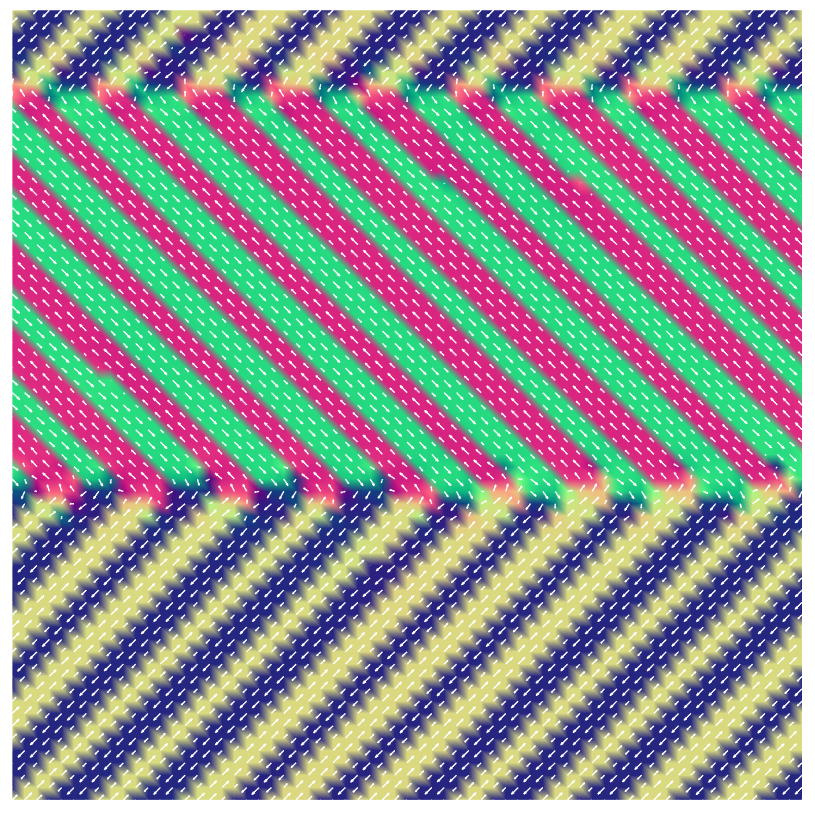} 
	} 
	\subfigure[ $n$=6] { \label{fig2:b} 
		\includegraphics[width=1.4in]{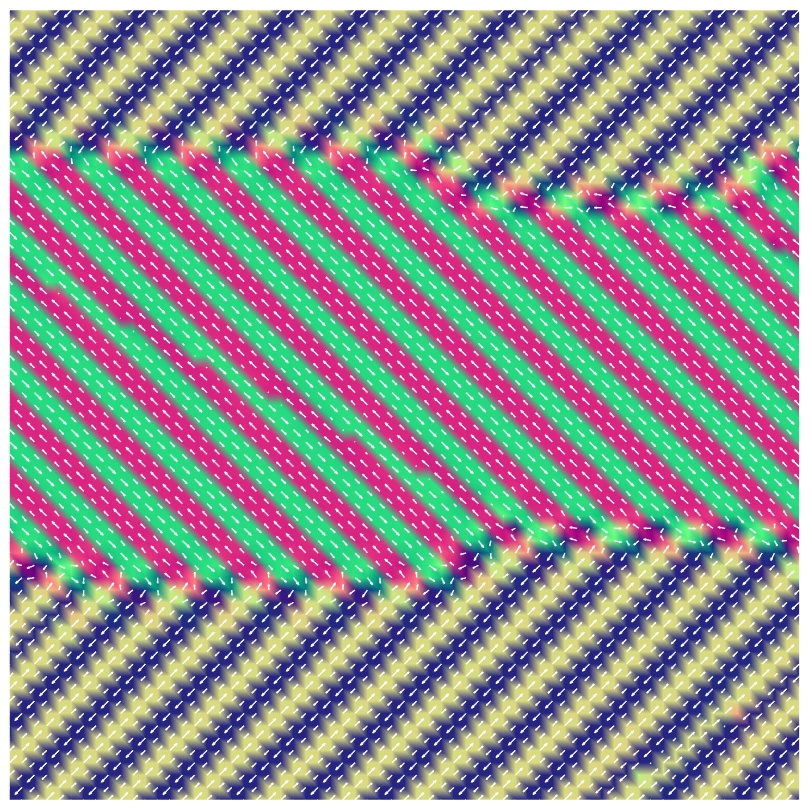} 
	} 
	\subfigure[ $n$=5] { \label{fig2:c} 
		\includegraphics[width=1.4in]{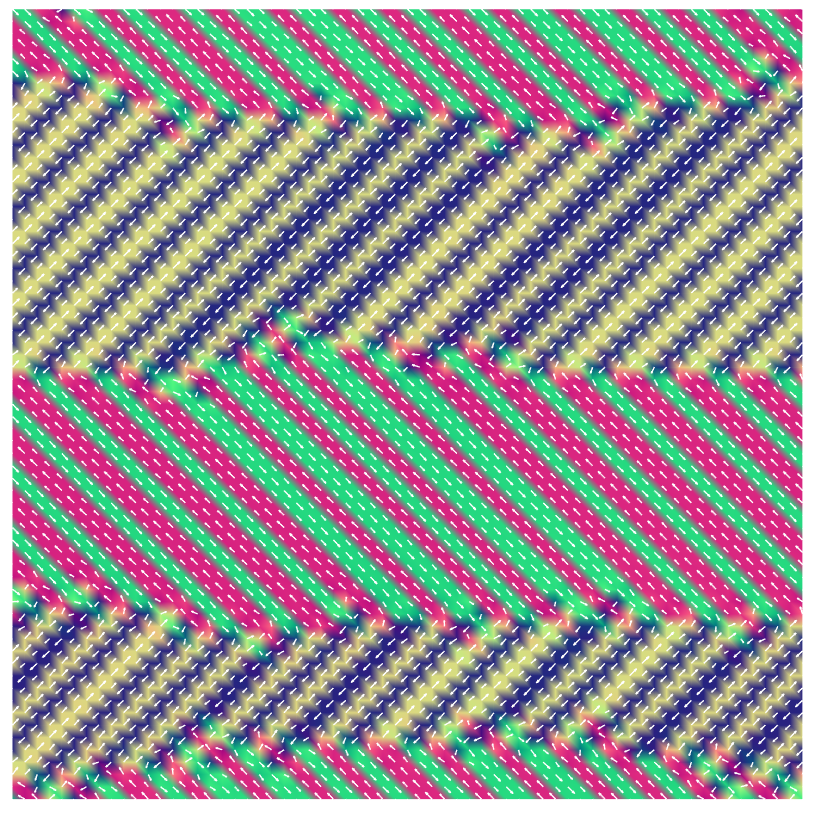} 
	} 
	\subfigure[ $n$=4] { \label{fig2:d} 
		\includegraphics[width=1.4in]{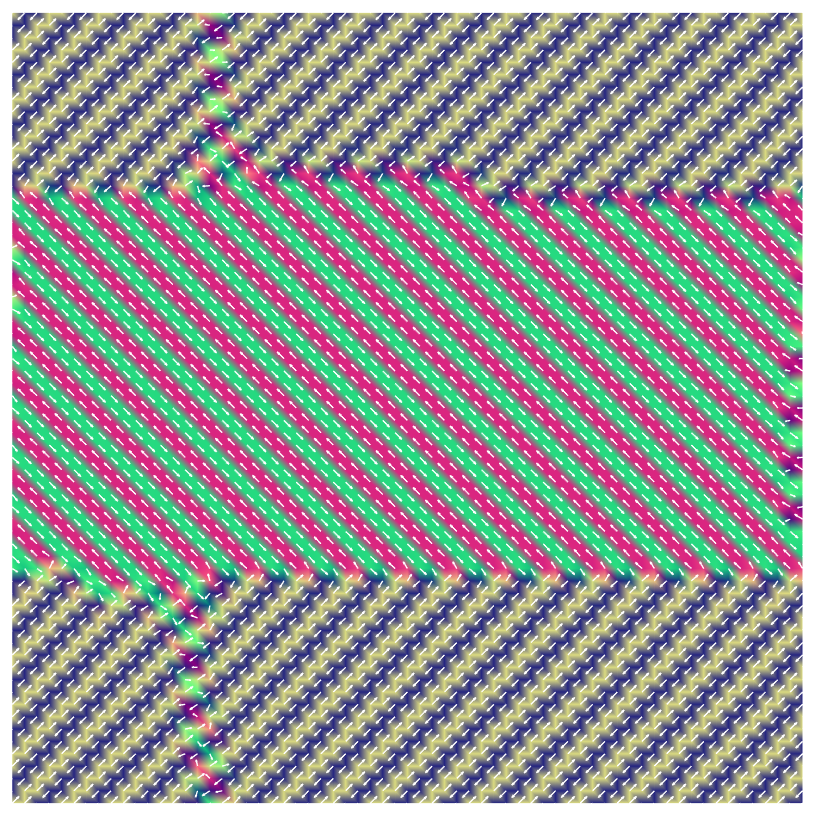} 
	} 
	\caption{The simulated phases with the increasing of the oxygen tilt angle. The white arrows represent the polarization vectors.  } 
	\label{fig2} 
\end{figure}

To investigate the domain structure and domain boundaries of AFE phases, we further carried out a phase field study based on the above theory. In the phase field study, besides the free energy $F$, the elastic energy $F_c$ and the electric static energy $F_e$ are also included in the total free energy density $H$, which is written as $H=F+F_c+F_e$. The elastic energy density is given by $F_c=\frac{1}{2}c_{ijkl}(\epsilon_{ij}-\epsilon_{ij}^0)(\epsilon_{kl}-\epsilon_{kl}^0)$, where $c_{ijkl}$ denotes the elastic constant, $\epsilon_{ij}$ is the total strain, and $\epsilon_{ij}^0$ represents the spontaneous strain. The spontaneous strain in terms of the polarization is given by $\epsilon_{ij}^0=Q_{ijkl}p_kp_l$, where $Q_{ijkl}$ is the electrostrictive coefficient. The solution of $F_c$ and $F_e$ can be found in the previous works\cite{liu2017enhanced, xu2009micromechanical, xu2010phase}. The temporal evolution of the polarization distribution is governed by the time-dependent Landau Ginzburg (TDLG) equation 
\begin{equation}
\frac{\partial p_i}{\partial t}=-L\frac{\delta H}{\delta p_i}
\end{equation}
where $L$ is the kinetic coefficient, and $t$ is the evolution time. Since we are interested in a qualitative understanding of the physical mechanisms, the Landau parameters are modified from the PZT system\cite{haun1989thermodynamic}, $\alpha_1=-5.54\times10^7$ JmC$^{-2}$, $\alpha_{11}=5.60\times10^8$ Jm$^{5}$C$^{-4}$, $\alpha_{111}=1.65\times10^9$Jm$^{9}$C$^{-6}$, $\alpha_{12}=2.89\times10^8$ Jm$^{5}$C$^{-4}$, $\alpha_{112}=-8.66\times10^8$ Jm$^{9}$C$^{-6}$, and $\alpha_{123}=3.19\times10^{10}$ Jm$^{9}$C$^{-6}$. The elastic constant $c_{11}=15.6\times10^{10}$ N/m$^2$, $c_{12}=9.6\times10^{10}$ N/m$^2$ and $c_{44}=12.7\times10^{10}$ N/m$^2$. And the electrostrictive coefficients are $Q_{11}=0.048$ m$^4$/C$^2$, $Q_{12}=-0.015$  m$^4$/C$^2$, and $Q_{44}=0.047$  m$^4$/C$^2$, respectively.
The oxygen tilt angle $\theta_1^2$ = $\theta_2^2$ = $\theta_3^2$ = $\theta^2$ due to soft R point mode.
For simplicity, a two dimensional system on a $64\times64$ grid with periodic boundary conditions is utilized in the simulations, the spacial step of each grid cell $\Delta x=a_c$, where $a_c=\frac{a_0}{\sqrt{2}}=0.416$ nm. We assume $\mu_{11}$, $\mu_{12}$, $\gamma_{11}$, and $\gamma_{12}$ to be zero, $g_{11}=g_{12}=1.25a_c^4\times10^7$ JmC$^{-2}$, and $\sigma_{11}$=$\sigma_{12}$=$\sigma_0'$ ($\sigma_0'>0$). Therefore, the value of $\sigma_0'\theta^2$ controls the modulation of the polarization phases. A small random fluctuation of polarization is employed as the initial condition.
\begin{figure}
	\centering
	\includegraphics[width=3.1in]{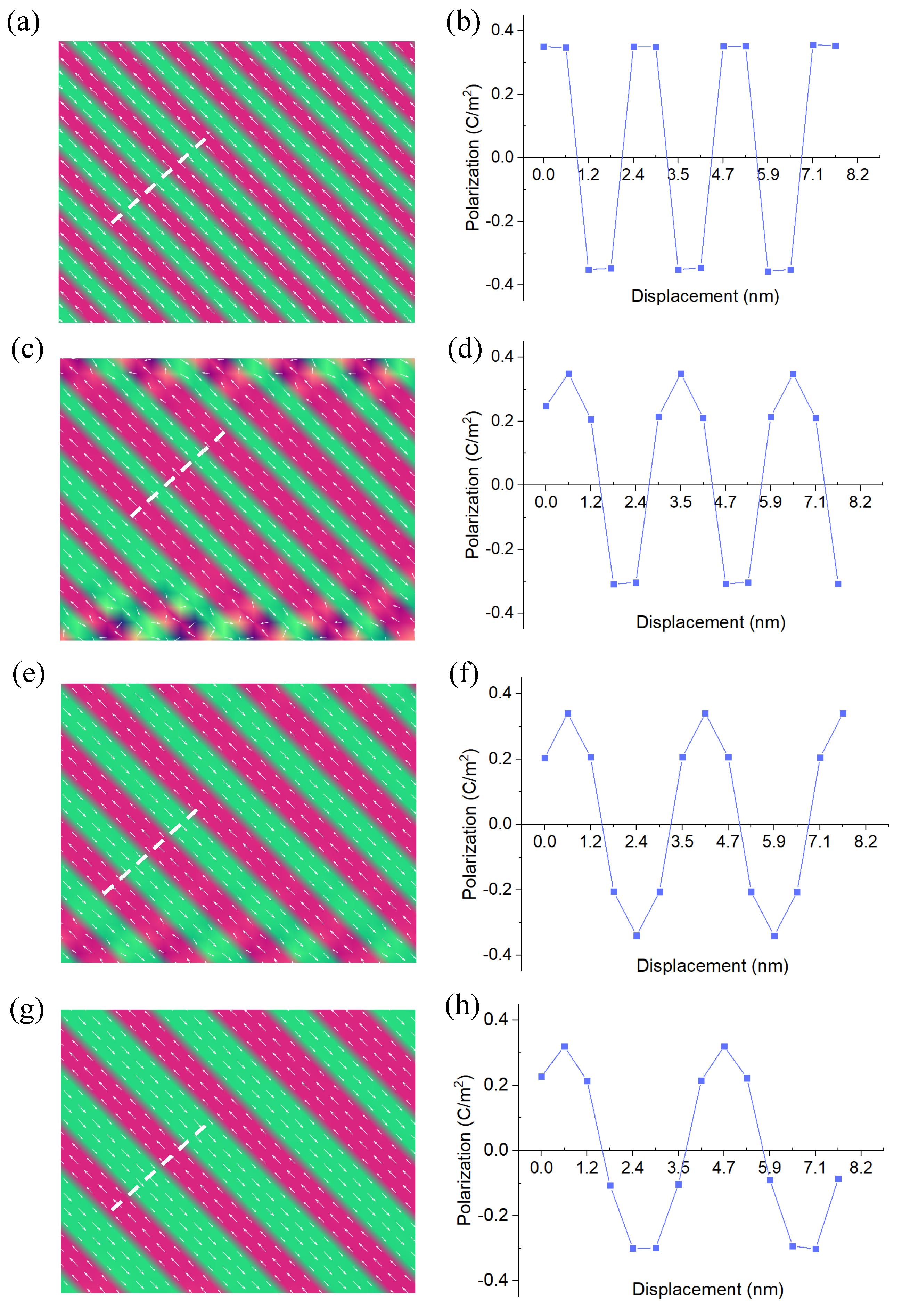}
	\caption{(a)-(d) The local polarization distribution in the phases with different $n$, the length of the white arrows represent the magnitude of the polarization. (e)-(h) The polarization magnitudes along the two antiparallel $\langle110\rangle_c$ directions related to the displacement along the length direction of the stripe domains.}
	\label{fig3}
\end{figure}

\begin{figure}
	\centering
	\includegraphics[width=3.2in]{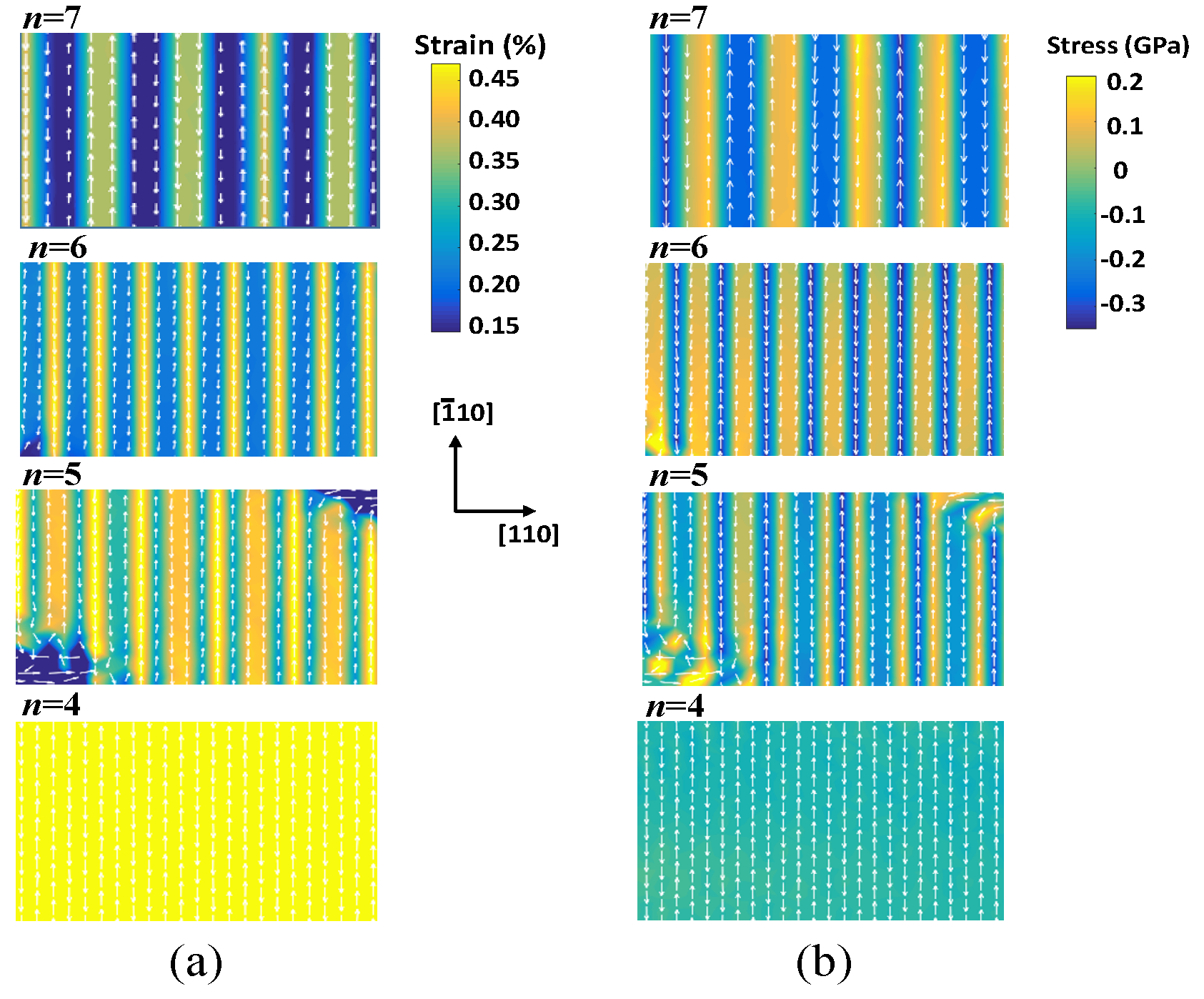}
	\caption{(a) Spontaneous strain of AFE phases along [$\bar{1}$10] direction . (b) Local stress field of AFE phases along [$\bar{1}$10] direction. The white arrows represent the polarization vectors.} 
	\label{fig4}
\end{figure}

Figure \ref{fig2} shows the simulated AFE structures by varying the value of $\sigma_0'\theta^2$. For $\sigma_0'\theta^2=1.75a_c^2\times10^7$ JmC$^{-2}$, the structure is AI phase ($\uparrow\uparrow\uparrow\downarrow\downarrow\downarrow\downarrow$) with $n=7$ in Figure \ref{fig2:a}. With the oxygen tilt increases, the periodicity $n$ decrease to 6 for $\sigma_0'\theta^2=2.31a_c^2\times10^7$ JmC$^{-2}$, resulting in a stable ($\uparrow\uparrow\uparrow\downarrow\downarrow\downarrow$) AC phase as shown in Figure \ref{fig2:b}. Further increasing of the $\sigma_0'\theta^2$ gives rise to a AI phase ($\uparrow\uparrow\downarrow\downarrow\downarrow$) with $n=5$, and it will transform to ($\uparrow\uparrow\downarrow\downarrow$) AFE phase when $\sigma_0'\theta^2$ increases to $3.12a_c^2\times10^7$ JmC$^{-2}$, as shown in Figure \ref{fig2:d}. Since the value of the oxygen tilt angle increases with the decrease of the temperature, the periodicity decreases with temperature cooling below the polarization transformation point. This transition behavior is widely observed in the PbZrO$_3$-based perovskite AFEs\cite{asada2004induced,watanabe2002features}.

Previously,  it was generally assumed that the AFE polarizations maintain the same magnitude across the domain boundaries and arrange fully compensated. However, the simulation results show that this understanding is not exactly correct. Figure \ref{fig3} (a)-(d) show the local polarization distribution of phases with different periodicity. The length of the white arrows represents the magnitude of the polarization. One can see that only the $n$=4 AFE phase has an arrangement of polarizations with the same magnitude, whereas for the other phases, the polarization is suppressed across the boundaries of AFE domains. The exact value of the polarization along the $\langle110\rangle_c$ direction of the corresponding phases are shown in Figure (e)-(h), in which the horizontal axis denotes the displacement along the length direction of the stripe domains. For $n=4$ and 6 AFE phases, the antiparallel polarizations are fully compensated, resulting in a zero total polarization. However, for AI phases with $n=5$ and 7, the polarization magnitudes of the two opposite directions are not equal, which gives rise to a remnant polarization. This unique behavior of AI phases from the simulation has recently been confirmed in the Pb-based perovskite AFEs\cite{ma2019uncompensated}. The spontaneous elastic strain of the local AFE domains along [$\bar{1}10$] direction  is calculated in Figure \ref{fig4}(a). One can see that the AFE lattices present lower spontaneous strain across the boundary in AFE phases with $n>4$. Whereas for $n=4$, the strain field is homogeneous. The variable strain field indicates that the AFE lattices can not be stress free. As shown in Figure \ref{fig4}(b), the calculated magnitude of the local stress can be as high as 0.15 GPa for the AFE phases with $n>4$. 

In conclusion, the transformation mechanism of AC and AI phases in perovskite AFEs is understood through the novel phenomenological model. Our phase field study shows that the polarization is suppressed across the AFE domain boundaries, giving rise to an inhomogeneous spontaneous elastic strain. Unlike the AC phases that form a fully compensated polarization arrangement, the AI phases usually present a remnant total polarization. The results also indicate that AFE states are not typically stress free, instead, there could be a high local stress field across the AFE domains. Our results lead to a new understanding of the morphology of AFE structures and domain boundaries.

\begin{acknowledgements}
This work was supported by the LOEWE program of the State of Hesse, Germany, within the project FLAME (Fermi Level Engineering of Antiferroelectric Materials for Energy Storage and Insulation Systems).
\end{acknowledgements}


\bibliography{reference}

\end{document}